\documentclass[aps,prl,twocolumn,showpacs,superscriptaddress,10pt]{revtex4-1}
\usepackage{amssymb,amsmath,units,times,graphicx,bm,textcomp}

\newcommand{\pSlash}{/\!\!\!p}
\newcommand{\kSlash}{/\!\!\!k}
\newcommand{\ASlash}{/\!\!\!\!A}
\newcommand{\epsSlash}{/\!\!\!\epsilon}

\newcommand{\e}{\text{e}}

\renewcommand{\d}{\text{d}}
\renewcommand{\i}{\text{i}}
\begin{document}

\title{Nonlinear double Compton scattering in the full quantum regime}

\author{F. Mackenroth}
\affiliation{Max-Planck-Institut f\"ur Kernphysik, Saupfercheckweg 1, 69117 Heidelberg, Germany}

\author{A. Di Piazza}
\email{dipiazza@mpi-hd.mpg.de}
\affiliation{Max-Planck-Institut f\"ur Kernphysik, Saupfercheckweg 1, 69117 Heidelberg, Germany}

\date{\today}
\begin{abstract}
A detailed analysis of the process of two photon emission by an electron scattered from a high-intensity laser pulse is presented. The calculations are performed in the framework of strong-field QED and include exactly the presence of the laser field, described as a plane wave. We investigate the full quantum regime of interaction, where photon recoil plays an essential role in the emission process, and substantially alters the emitted photon spectra as compared to those in previously-studied regimes. We provide a semiclassical explanation for such differences, based on the possibility of assigning a trajectory to the electron in the laser field before and after each quantum photon emission. Our numerical results indicate the feasibility of investigating experimentally the full quantum regime of nonlinear double Compton scattering with already available plasma-based electron accelerator and laser technology.
\end{abstract}

\maketitle

Electromagnetic radiation by accelerated charged particles is one of the most fundamental processes in physics and it is exploited experimentally for different purposes, spanning from the generation of coherent x-ray fields \cite{Moshammer_2009} and of even multi-GeV photon beams \cite{Apyan_2005} to medical applications. From a more fundamental point of view and by limiting to the case of electromagnetic driving fields, the process of electromagnetic radiation by accelerated charges has played a crucial role for testing the validity of classical electrodynamics and of QED. In QED the emission of radiation by a charge, an electron for definiteness (mass $m$ and charge $e<0$, respectively), is described as the emission of quanta of the electromagnetic field, i.e., of photons. As in classical electrodynamics an electron emits only if it is accelerated, quantum mechanically photon emission can occur only if the electron absorbs at least one photon. (Linear) Compton scattering, i.e., the emission of a single photon by an electron via the absorption of another photon, has been investigated theoretically and experimentally since the formulation of QED itself, and it has provided relevant information on the structure of the theory at high energies \cite{Achard_2005}. The related process in which two photons are emitted ((linear) double Compton scattering) has also been investigated both theoretically \cite{PertDoubleCompton_Theo} and experimentally \cite{PertDoubleCompton_Exp}.

When an electron is driven by an intense electromagnetic field, the emission of photons may occur with the absorption of many photons from the field. High-power lasers are sources of strong electromagnetic fields with unprecedented intensities \cite{Di_Piazza_2012} and they represent an irreplaceable tool to test the high-intensity or strong-field sector of QED, as complementary to the high-energy one. In strong-field QED the background field is so intense that it has to be taken into account exactly in the calculations. For a laser field approximated as a plane wave, this is required when the Lorentz- and gauge-invariant parameter $\xi=|e|\mathcal{E}_L/mc\omega_L$ is of order of or larger than unity, where $\mathcal{E}_L$ is the electric field amplitude of the wave and $\omega_L$ its central angular frequency. The threshold $\xi\approx 1$ corresponds to an optical ($\hbar\omega_L \approx 1\;\text{eV}$) laser intensity of about $10^{18}\;\text{W/cm$^2$}$, which is four orders of magnitude smaller than available optical laser intensities \cite{Yanovsky2008}. The emission of a single photon by an electron in the field of a plane wave (nonlinear single Compton scattering (NSCS)) has been investigated theoretically since the invention of the laser (see the recent review \cite{Di_Piazza_2012}) and a pioneering experiment performed at SLAC \cite{NonlinearQR_Exp} has confirmed the predictions of QED at $\xi\lesssim 1$. More recent studies on NSCS have been focused especially on the high-intensity regime $\xi\gg 1$, where a large number ($\sim \xi^3$) of laser photons is absorbed by the electron during the emission process, and on finite-pulse effects \cite{Boca2009,Mackenroth2011,Seipt2011,Krajewska2012}. NSCS is found to be characterized also by the so-called quantum nonlinearity parameter $\chi=((k_Lp)/m\omega_L)(\mathcal{E}_L/\mathcal{E}_{\text{cr}})$. Here, $k_L^{\mu}=(\omega_L/c,\bm{k}_L)$ is the four-wavevector of the laser photons ($|\bm{k}_L|=\omega_L/c$), $p^{\mu}=(\varepsilon/c,\bm{p})$ is the initial four-momentum of the electron and $\mathcal{E}_{\text{cr}}=m^2c^3/\hbar|e|=1.3\times 10^{16}\;\text{V/cm}$ is the critical field of QED \cite{Di_Piazza_2012}. The parameter $\chi$ controls quantum effects like the photon recoil and at $\chi\ll 1$ the NSCS spectra coincide with the classical ones \cite{Jackson}. The emission of two photons by an electron in the field of a plane wave (nonlinear double Compton scattering (NDCS)) has also been studied in the literature, with an emphasis on the correlation of the two emitted photons \cite{Lotstedt2009a} and on the relative yield between NSCS and NDCS \cite{Seipt2012}. Both these studies have been focused on the radiation regime where $\xi\sim 1$ such that the electron absorbs only a few photons from the laser field (``quasilinear'' regime), and where $\chi\ll 1$, such that quantum photon recoil was negligible.

In the present Letter we investigate NDCS in the full non-perturbative quantum regime where $\xi\gg 1$ and $\chi\gtrsim 1$. By including exactly the effects of the plane wave, we show that the emission spectra in the full quantum regime substantially differ from those at $\chi\ll 1$ as a result of the photon recoil. We explain the new features of the emission spectra, by developing a quasiclassical approach, based on the possibility of assigning a trajectory to the electron in the laser field, with discontinuities in the electron energy due to the quantum emission of photons. Finally, by means of a numerical simulation, we show that the full quantum regime can be entered already with experimentally demonstrated laser intensities of the order of $10^{21}\, \text{W}/\text{cm}^2$ \cite{Yanovsky2008} together with presently available plasma-based electron accelerator technology \cite{ElectronEnergies}.

\begin{figure}
\centering
\includegraphics[width=.6\linewidth]{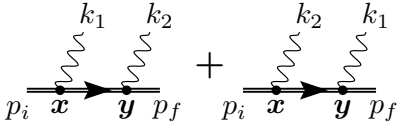}
\caption{Tree-level Feynman diagrams of NDCS in the Furry picture. The double solid lines represent Volkov states and propagators.}
\label{fig:NDCS_FeynDiags}
\end{figure}

We consider a linearly-polarized plane-wave field, described by the four-potential $A_L^{\mu}(\eta)=(\mathcal{E}_L/\omega_L)\epsilon_L^{\mu}\psi_{\mathcal{A}}(\eta)$, where $\epsilon^{\mu}_L$ is the wave's polarization four-vector, and $\psi_{\mathcal{A}}(\eta)$ gives the field's temporal shape, which depends on the space-time coordinates $x^{\mu}$ only via the invariant phase $\eta=(k_Lx)$ (units with $\hbar = c = 1$ are used throughout). For a generic four-vector $a^{\mu}=(a^0,\bm{a})$ it is convenient to introduce the light-cone representation $a^{\mu}=(a^+,a^{\text{\textminus}},\bm{a}^{\perp})$, where $a^{\pm}=\left(a^0\pm a^{\|}\right)/\sqrt{2}$, with $a^{\|}=\bm{k}_L\cdot\bm{a}/\omega_L$, and where $\bm{a}^{\perp}=\bm{a}-a^{\|}\bm{k}_L/\omega_L$. In the Furry picture of QED, the electron wave functions are solutions of the Dirac equation in the presence of the considered background field, i.e., Volkov wave functions in the case 
of a plane-wave field \cite{Landau_b_4_1982}. The Volkov wave function for an electron with four-momentum $p^{\mu}$ and spin quantum number $\sigma$ outside the plane wave has the form $\Psi_{p,\sigma}(x)=E_p(x)u_{p,\sigma}/\sqrt{2\varepsilon}$, where $u_{p,\sigma}$ is a free bi-spinor, where a unity quantization volume is assumed, and where the Ritus matrices
\begin{equation}
E_p(x)=\left[1+ \frac{e\kSlash_L\ASlash_L(\eta)}{2(k_Lp)}\right]e^{-\i\left\{px+\int_0^{\eta}\d\phi \left[\frac{e (pA_L(\phi))}{(k_Lp)}-\frac{e^2A_L^2(\phi)}{2(k_Lp)}\right]\right\}}
\end{equation}
have been introduced, with the notation $/\!\!\!a=\gamma^{\mu}a_\mu$ for the Dirac matrices $\gamma^\mu$. We consider an electron with initial (final) four-momentum $p_i^{\mu}=(\varepsilon_i,\bm{p}_i)$ ($p_f^{\mu}=(\varepsilon_f,\bm{p}_f)$) and spin quantum number $\sigma_i$ ($\sigma_f$), which emits two photons with four-momenta $k^{\mu}_1$ and $k_2^{\mu}$ and with polarization four-vectors $\epsilon^{\mu}_{k_1,\lambda_1}$ and $\epsilon^{\mu}_{k_2,\lambda_2}$, respectively. The scattering matrix element $S_{fi}$ of this process can be written as $S^{(1)}_{fi}+S^{(2)}_{fi}$, where (see Fig.\ \ref{fig:NDCS_FeynDiags})
\begin{align}
S^{(1)}_{fi} =& - e^2\int \d^4 x \d^4 y\ \overline{\Psi}_{p_f,\sigma_f}(y)\,\epsSlash_{k_2,\lambda_2}^*\e^{\i k_2y}\,G(y,x)\,\epsSlash_{k_1,\lambda_1}^*\nonumber \\
&\times\e^{\i k_1x}\,\Psi_{p_i,\sigma_i}(x). \label{Eq:Matrix_Element_int}
\end{align}
and $S^{(2)}_{fi}$ is obtained from $S^{(1)}_{fi}$ by exchanging the photon indices $1$ and $2$. In Eq. (\ref{Eq:Matrix_Element_int}) we have introduced $\overline{\Psi}_{p,\sigma}(x)=\Psi^{\dag}_{p,\sigma}(x)\gamma^0$, and the dressed electron propagator \cite{Ritus1985}
\begin{equation}
 G(y,x) = \lim_{\epsilon\to0}\int \frac{\d^4p}{(2\pi)^4}E_p(y)\frac{\pSlash+m}{p^2-m^2+\i \epsilon}\overline{E}_p(x) \label{Eq:Propagator}
\end{equation}
with $\overline{E}_p(x)=\gamma^0E^{\dag}_p(x)\gamma^0$. It is clear that it is sufficient to analyze here only the quantity $S^{(1)}_{fi}$. The structure of the Ritus matrices allows to perform the integrals with respect to the coordinates $x^+$ and $\bm{x}^{\perp}$ ($y^+$ and $\bm{y}^{\perp}$), which provide the energy-momentum conservation laws $\bm{p}_i^{\perp}=\bm{k}_1^{\perp}+\bm{p}^{\perp}$ and $p_i^-=k_1^-+p^-$ ( $\bm{p}^{\perp}=\bm{k}_2^{\perp}+\bm{p}_f^{\perp}$ and $p^-=k_2^-+p_f^-$) at vertex $x$ ($y$). One of these sets of conservation laws can be employed to perform three integrals in the electron propagator. The only remaining integral $\mathcal{I}(y^-,x^-)$ in Eq.\ (\ref{Eq:Propagator}) is
\begin{align}
\mathcal{I}(y^-,x^-) &= \lim_{\epsilon\to0}\int \frac{\d p^+}{2\pi}\frac{\pSlash+m}{p^+-p^+_t+\i \epsilon}\e^{\i p^+(x^--y^-)}\nonumber \\
 &=\not \!n_L\delta(x^{\text{\textminus}}-y^{\text{\textminus}})-\i (\pSlash_t+m)\Theta(y^{\text{\textminus}}-x^{\text{\textminus}}), \label{Eq:PropagatorDecomposition}
\end{align}
where $n_L^{\mu}=k_L^{\mu}/\omega_L$, $\Theta(\cdot)$ is the step function, and where we have introduced the ``transitional'' four-momentum $p_t^{\mu}$ with light-cone coordinates $p_t^-=p_i^--k_1^-=k_2^-+p_f^-$, $\bm{p}_t^{\perp}=\bm{p}_i^{\perp}-\bm{k}_1^{\perp}=\bm{k}_2^{\perp}+\bm{p}_f^{\perp}$ and $p_t^+=({\bm{p}_t^{\perp}}^2+m^2)/2p_t^-$. Note that the four-momentum $p_t^{\mu}$ is on-shell, i.e., it fulfills the condition $p_t^2=m^2$. The above decomposition allows to write the quantity $S^{(1)}_{fi}$ as $S^{(1)}_{fi}=(2\pi)^3\delta(p_i^--k_1^--k_2^--p_f^-)\delta^{(2)}(\bm{p}_i^{\perp}-\bm{k}_1^{\perp}-\bm{k}_2^{\perp}-\bm{p}_f^{\perp})\sum_{r,s=0}^2(a_rf_r\delta_{r,s}+b_{r,s}f_{r,s})$, where the coefficients $a_r$ and $b_{r,s}$ are matrix factors given by products of a large number of Dirac matrices, whose exact form is not needed here. In fact, all the dynamical information on the process is contained in the functions
\begin{subequations}\label{Eq:DynamicIntegrals}
 \begin{align}
  f_r =& \int \d \eta \psi_{\mathcal{A}}^r(\eta)\ \text{exp}\{-\i\,[S_x(\eta)+S_y(\eta)]\}, \label{Eq:DynamicIntegrals_fi}\\
  f_{r,s} =& \int \d \eta_x\d \eta_y \Theta(\eta_y-\eta_x)\ \psi_{\mathcal{A}}^s(\eta_x)\psi_{\mathcal{A}}^r(\eta_y) \nonumber \\
  &\times\text{exp}\{-\i\,[S_x(\eta_x)+S_y(\eta_y)]\}, \label{Eq:DynamicIntegrals_fij}
 \end{align}
\end{subequations}
where $S_{x/y}(\eta)=\int_0^{\eta}\d \eta'[\alpha_{x/y}\,\psi_{\mathcal{A}}(\eta')+\beta_{x/y}\,\psi_{\mathcal{A}}^2(\eta')+\gamma_{x/y}]$, with $\alpha_x = -m\xi[(p_i\epsilon_L)/(k_Lp_i)-(p_t\epsilon_L)/(k_Lp_t)]$, $\beta_x = -m^2\xi^2(k_Lk_1)/2(k_Lp_t)(k_Lp_i)$, $\gamma_x = -(k_1p_i)/(k_Lp_t)$ and with $\alpha_y$, $\beta_y$ and $\gamma_y$ obtained from $\alpha_x$, $\beta_x$ and $\gamma_x$, respectively, with the substitutions $p^{\mu}_t\to p^{\mu}_f$, $p^{\mu}_i\to p^{\mu}_t$ and $k^{\mu}_1\to k^{\mu}_2$. The integrals, which do not contain the shape function $\psi_{\mathcal{A}}(\eta)$ in the prefactor, diverge. These divergences can be avoided by employing the identities \cite{Mackenroth2011,Seipt2012,Ilderton2011} $\gamma_y\,f_{0,j} = -\i f_j-\alpha_y\,f_{1,j}+\beta_y\,f_{2,j}$, $\gamma_x\,f_{i,0} = \i f_i-\alpha_x\,f_{i,1}+\beta_x\,f_{i,2}$ and $(\gamma_x+\gamma_y) f_0 = -(\alpha_x+\alpha_y)f_1-(\beta_x+\beta_y)f_2$. The differential average energy emitted $\d E$, summed (averaged) over all outgoing (incoming) discrete quantum numbers is given by
\begin{equation}
\d E = \frac{\omega_1+\omega_2}{2}\frac{d^3\bm{p}_f}{(2\pi)^3}\prod_{i=1}^2\frac{d^3\bm{k}_i}{(2\pi)^3}\sum_{\{\sigma,\lambda\}}\Big|S^{(1)}_{fi}+S^{(2)}_{fi}\Big|^2 ,
\end{equation}
where $\{\sigma,\lambda\}\equiv \sigma_i,\sigma_f,\lambda_1,\lambda_2$. Note that the three $\delta$-functions contained in $S^{(1)}_{fi}$ (and equally in $S^{(2)}_{fi}$) can be exploited to perform the integrals in $\bm{p}_f$. 

We will consider the electron-laser collision in a reference frame where the laser pulse propagates along the positive $z$-axis ($n^{\mu}_L=(1,0,0,1)$), it is polarized along the $x$-direction ($\epsilon_L^{\mu}=(0,1,0,0)$) and the electron is initially counterpropagating with respect to the laser beam, i.e., $\bm{p}_i=(0,0,-\beta_i \varepsilon_i)$, with the electron's initial velocity $-\beta_i<0$. In order to stress differences to existing treatments valid for monochromatic fields \cite{Lotstedt2009a}, the laser's temporal pulse shape is modeled by the function $\psi_{\mathcal{A}}(\eta)=\sin^4(\eta/4)\sin(\eta)$ for $\eta\in\left[0,4\pi\right]$ and zero elsewhere, corresponding to a two-cycle pulse of approximately $5$-fs duration at $\omega_L=1.55\;\text{eV}$. We first consider a laser system with peak intensity $I_L=5\times 10^{20}\; \text{W}/\text{cm}^2$, corresponding to a classical nonlinearity parameter $\xi\approx15$. In this regime the process exhibits a highly nonlinear dependence on the laser amplitude, as about $\xi^3\sim 3000$ photons are estimated to be absorbed by the electron from the laser field \cite{Ritus1985,Mackenroth2011}. Moreover, in order to highlight the qualitative differences between the full quantum regime $\chi\approx 1$ considered here and the already-studied regime at $\chi\ll 1$, we first consider a numerical example within the latter. Thus, we set $\varepsilon_i=40$ MeV, which corresponds to $\chi=5\times 10^{-3}$, ensuring that photon-recoil effects are negligible. We choose to observe one photon at $\theta_1=\pi-\theta_0/2$, with $\theta_0=m\xi/\varepsilon_i$, and the other one at the two different polar angles $\theta_2=\theta_1$ (see Fig. 2a) and  $\theta_2=\pi-1.1\,\theta_0$ (see Fig. 2b). Also, we choose $\phi_1=\pi,\phi_2=0$ in both cases as azimuthal observation angles. We observe that for our pulse shape $\psi_{\mathcal{A}}(\eta)$: 1) the emission cone of NSCS is determined by the condition $\pi-\theta\le 0.8\,\theta_0$ on the polar angle $\theta$ (the prefactor of $\theta_0$ is given by the numerical maximum of the function $\psi_{\mathcal{A}}(\eta)$) \cite{Mackenroth2010}, and 2) the NSCS emission spectra at $(\theta,\phi=0)$ and at $(\theta,\phi=\pi)$ coincide. Therefore, if $\theta_2=\theta_1$ both photons are observed within the emission cone of NSCS, whereas if $\theta_2=\pi-1.1\,\theta_0$ one of the photons is observed outside this cone.
\begin{figure}
\includegraphics[width=\linewidth]{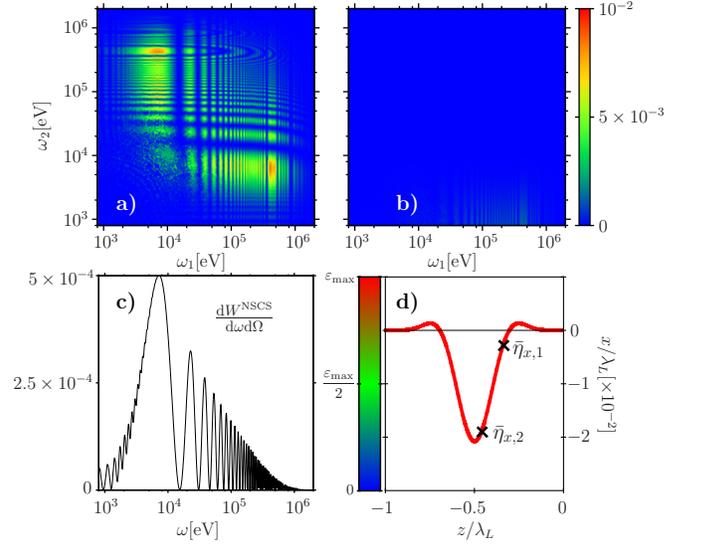}
\caption{(color online) Two-photon energy emission spectra $\d E/\Pi_{i=1}^2\d \omega_i\d\Omega_i[\text{eV$^{-1}$ sr$^{-2}$}]$ at $\chi= 5\times 10^{-3}$ observed at $\theta_1=\pi-\theta_0/2$, and at $\theta_2=\theta_1$ (part a)) and at $\theta_2=\pi-1.1\,\theta_0$ (part b)), with $\theta_0=0.19\;\text{rad}$. The other numerical parameters are given in the text. Part c): NSCS emission probability $\d W^{\text{NSCS}}/\d\omega\d\Omega[\text{eV$^{-1}$ sr$^{-1}$}]$ at ($\theta_1$,$\phi_1$). Part d): classical electron trajectories (the instantaneous electron energy is color-encoded) with initial four-momenta $p^{\mu}_i$ and $p^{\mu}_t$, joined at $\bar{\eta}_{x,1}$. Since as a typical photon emission energy the value $10^5\;\text{eV}$ has been chosen (see parts a) and c)), then $p^{\mu}_i\approx p^{\mu}_t$ and the two trajectories are indistinguishable. $\varepsilon_{\text{max}}$ is the maximum electron energy in the laser field. The crosses mark the points of the trajectory, where the classical electron's velocity is along ($\theta_1$,$\phi_1$).}
 \label{fig:NoRecoil_Spectra}
\end{figure}
Figures 2a and 2b show that there is emission of radiation only within the NSCS emission cone. This feature can be quantitatively understood by virtue of a stationary-phase analysis \cite{Ritus1985,Mackenroth2011}. In fact, in the regime $m\xi,\varepsilon_i\gg m$ the phases in the integrands in Eqs.\ (\ref{Eq:DynamicIntegrals}) are of the order of $\xi^3$ and thus very large \cite{Ritus1985,Mackenroth2011}, and the saddle-point method can be applied to evaluate the functions $f_r$ and $f_{r,s}$. In the case of NSCS, the stationary points for a given observation direction correspond to those phase instants where the classical electron velocity points towards that observation direction \cite{Mackenroth2011}. Accordingly, the overall emission cone corresponds to the angular region that is spanned by the classical electron's velocity vector along the complete classical electron's trajectory in the laser pulse. Thus, the electron propagation in the laser field is quasiclassical and the photon recoil is the main quantum effect to be accounted for \cite{Baier_b_1998}. Now, we have checked numerically that the emission spectrum is dominated by the contribution proportional to the functions $f_{r,s}$. These integrals, according to the stationary phase method, can be approximated as $f_{r,s}\approx\sum_{l,n}\Theta(\bar{\eta}_{y,n}-\bar{\eta}_{x,l})f^y_r(\bar{\eta}_{y,n})f^x_s(\bar{\eta}_{x,l})$. Here, the indices $l$ and $n$ run over all stationary points, which are found as solutions of the equations $dS_{x/y}(\eta)/d\eta\vert_{\eta=\bar{\eta}_{x,l/y,n}}=0$ and $f^{x/y}_r = \int \d \eta \psi_{\mathcal{A}}^r(\eta)\ \text{exp}[-\i S_{x/y}(\eta)]$ (see Eq. (\ref{Eq:DynamicIntegrals_fij})). The above-approximated expression of $f_{r,s}$ can be interpreted as a two-step emission process in which the electron first emits a photon with four-momentum $k^{\mu}_1$ changing its own four-momentum from the initial one $p^{\mu}_i$ to the transitional one $p^{\mu}_t$, and then it emits a photon with four-momentum $k^{\mu}_2$ changing its own four-momentum from the transitional one $p^{\mu}_t$ to the final one $p^{\mu}_f$ (recall that the total amplitude also contains a term with the photon indices 1 and 2 exchanged). In addition, one can picture this dynamics as a succession of two classical trajectories with initial four-momenta $p^{\mu}_i$ and $p^{\mu}_t$, respectively, continuously joined at a point corresponding to the phase $\bar{\eta}_{x,l}$ where the photon with four-momentum $k^{\mu}_1$ is emitted. In Fig.\ \ref{fig:NoRecoil_Spectra}d we show in principle a pair of such classical electron trajectories joined at the point marked with a cross and labeled as $\bar{\eta}_{x,1}$ where the electron propagates along the observation direction ($\theta_1,\phi_1$). However, since in the regime $\chi\ll 1$ the recoil is negligible (see Figs.\ \ref{fig:NoRecoil_Spectra}a and 2c and recall that $\varepsilon_i=40\;\text{MeV}$), the two four-momenta $p^{\mu}_i$ and $p^{\mu}_t$ are practically identical and the two trajectories are indistinguishable (in Fig.\ \ref{fig:NoRecoil_Spectra}d we subtracted the momentum of a typical photon of energy $\omega_1=10^5\;\text{eV}$ emitted towards ($\theta_1,\phi_1$) from the initial electron momentum, see Figs.\ \ref{fig:NoRecoil_Spectra}a and 2c). The same occurs if the two trajectories are joined at the other saddle point $\bar{\eta}_{x,2}$ (see Fig.\ \ref{fig:NoRecoil_Spectra}d), where the electron's velocity is again along the observation direction ($\theta_1,\phi_1$). From an analytical point of view, by inserting the parameters $\alpha_x$ and $\beta_x$ ($\alpha_y$ and $\beta_y$), the stationary-point equation at the vertex $x$ ($y$) evaluates to the approximate conditions $\psi_{\mathcal{A}}\left(\bar{\eta}_x\right) = -1/2$ ($\psi_{\mathcal{A}}(\bar{\eta}_y) = \Delta\vartheta_2 - (\omega_1/\varepsilon_i)(1/2 + \Delta\vartheta_2)$, with $\Delta\vartheta_2=(\pi-\theta_2)/\theta_0$). If photon recoil is negligible ($\omega_1\ll\varepsilon_i$), we observe that the condition for $\bar{\eta}_y$ has a real solution only if $\Delta\vartheta_2<0.8$, i.e., only if $\theta_2$ lies within the NSCS emission cone. This explains the absence of emission outside this cone in Fig.\ \ref{fig:NoRecoil_Spectra}b. Also, by comparing now the frequency distribution of the NDCS emission spectrum in Fig.\ \ref{fig:NoRecoil_Spectra}a with the NSCS emission probability $\d W^{\text{NSCS}}/\d\omega\d\Omega$ \cite{Boca2009,Mackenroth2011,Seipt2011} in Fig.\ \ref{fig:NoRecoil_Spectra}c, it is apparent that the NDCS spectrum for $\chi\ll1$ corresponds to an emission probability given by the ``product'' of two independent NSCS probability distributions for each photon (note that the maximum in the probability in Fig. 2c is at lower energies than in the two-photon emission spectrum in Fig. 2a, as the latter contains an additional factor $\omega_1+\omega_2$). Since the formation length of NSCS at $\xi\gg 1$ is much smaller than the laser's central wavelength $\lambda_L=2\pi/\omega_L$ \cite{Di_Piazza_2012}, if recoil effects can be neglected, the two-photon emission process can be described by two independent NSCS events \cite{Glauber_1951}. Accordingly, we checked that if $W^{\text{NDCS}}$ ($W^{\text{NSCS}}$) is the total NDCS (NSCS) emission probability, then $W^{\text{NDCS}}\approx (W^{\text{NSCS}})^2/2$ \cite{Glauber_1951}.

\begin{figure}
\includegraphics[width=\linewidth]{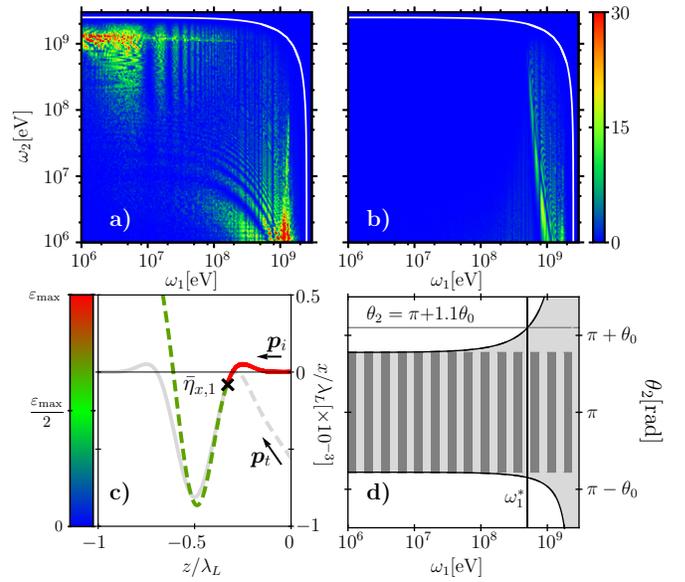}
\caption{(color online) Two-photon energy emission spectra $\d E/\Pi_{i=1}^2\d \omega_i\d\Omega_i [\text{eV$^{-1}$ sr$^{-2}$}]$ at $\chi\approx 1.1$ observed at $\theta_1=\pi-\theta_0/2$, and at $\theta_2=\theta_1$ (part a)) and at $\theta_2=\pi-1.1\,\theta_0$ (part b)), with $\theta_0=7.6\times 10^{-3}\;\text{rad}$. The other numerical parameters are given in the text. The solid white lines correspond to the cutoff-energy equation $\omega_1+\omega_2=\varepsilon_i$. Part c) the two classical electron trajectories with initial electron momentum $\bm{p}_i$ (solid line) and $\bm{p}_t$ (dashed line). The color-encoded line shows the actual electron trajectory for a photon with energy $\omega_1=0.8$ GeV and momentum along $(\theta_1,\phi_1)$ emitted at $\bar{\eta}_{x,1}$. Part d) Emission opening angle for the second emitted photon as a function of $\omega_1$ (light shaded area) compared with the emission cone for NSCS with initial electron momentum $\bm{p}_i$ (dark stripes). The vertical line indicates the value $\omega_1=\omega_1^*$ described in the text.}
 \label{fig:Recoil_Spectra}
\end{figure}

The physical situation, however, changes substantially if we enter the full quantum regime at $\chi \approx 1$. In order to investigate this regime, we set $\varepsilon_i=2.5$ GeV, as was deemed achievable by current technology \cite{Kalmykov2010}, and $I_L=3\times 10^{21} \text{W}/\text{cm}^2$ ($\xi\approx37$), resulting in $\chi=1.1$, and keep all other parameters unchanged with respect to the above example. By observing the two emitted photons at the same emission angles as before (see Figs.\ \ref{fig:Recoil_Spectra}a and \ref{fig:Recoil_Spectra}b),
we note that: 1) the quantum mechanical cutoff energy for the sum of the emitted photons' energies, approximately given by the equation $\omega_1+\omega_2=\varepsilon_i$, is well approached; 2) since in the quantum regime the electron loses a substantial part of its energy after the emission of the first photon, the asymmetry in the energies of the two emitted photons (see Fig.\ \ref{fig:Recoil_Spectra}a) is much more pronounced than at $\chi\ll 1$ (see Fig.\ \ref{fig:NoRecoil_Spectra}a); 3) the electron also emits outside of the NSCS emission cone (see Fig.\ \ref{fig:Recoil_Spectra}b and note that $\theta_0=7.6\times 10^{-3}\;\text{rad}$ for the present numerical parameters). This last feature is particularly important as it can be exploited to measure NDCS in the full quantum regime. In order to explain it qualitatively, we show in Fig.\ \ref{fig:Recoil_Spectra}c the classical trajectories for initial momenta $\bm{p}_i$ (solid line) and $\bm{p}_t$ (dashed line), obtained for the emission of a photon of energy $\omega_1=0.8$ GeV and momentum  along $(\theta_1,\phi_1)$ at $\bar{\eta}_{x,1}$ (the analysis in the case in which the electron emits at $\bar{\eta}_{x,2}$ is analogous). At this value of $\omega_1$ the emission probability is maximal. The derivative of the two trajectories is continuous at $\bar{\eta}_{x,1}$ (see Fig.\ \ref{fig:Recoil_Spectra}c), as the photon is assumed to be emitted with momentum parallel to the instantaneous electron's velocity. However, the color-coding shows that the electron energy discontinuously decreases at the emission point due to photon recoil. The abrupt decrease in energy induces a stronger deflection of the electron trajectory in the laser field after the photon emission. Thus, the resulting emission cone's opening angle increases and the emission outside of the NSCS emission cone becomes possible. The same conclusion can be drawn analytically, based on evaluating the saddle-point equation $\psi_{\mathcal{A}}(\bar{\eta}_y) = \Delta\vartheta_2 - (\omega_1/\varepsilon_i)(1/2 + \Delta\vartheta_2)\le 0.8$, with $\Delta\vartheta_2=(\pi-\theta_2)/\theta_0$. In fact, solving the latter inequality for $\theta_2$, we obtain an expression for the cutoff angles of the second photon emission as a function of $\omega_1$ (see Fig.\ \ref{fig:Recoil_Spectra}d, with $\phi_2=0$ mapped to $\theta_2>\pi$ and $\phi_2=\pi$ mapped to $\theta_2<\pi$). The photon-energy threshold $\omega_1^*$, beyond which emission along $\theta_2=\pi+1.1\,\theta_0$ becomes possible (see Fig. \ref{fig:Recoil_Spectra}b) is well reproduced by this analytical prediction of the angular emission region (see Fig. \ref{fig:Recoil_Spectra}d). In this way, measuring MeV photons outside of the NSCS emission cone would reveal a NDCS signal, where NSCS is exponentially suppressed and negligible. We have estimated numerically that the emission probability outside of the NSCS emission cone is of the order of 0.1, indicating the observability in principle of the process, employing laser-generated electron beams, which typically contain of the order of $10^8$ electrons \cite{ElectronEnergies}.

In summary, we have analyzed for the first time nonlinear double Compton scattering in a plane-wave field in the full non-perturbative quantum regime, where a large number of laser photons are absorbed by the electron during the emission process, and where photon recoil substantially alters the electron's dynamics. This alteration is, in particular, responsible for the emission of photons along directions, where nonlinear single Compton scattering is suppressed, which provides a simple way of detecting double Compton scattering in the full quantum regime already with available laser and electron acceleration technology.

The authors acknowledge useful discussions with A. Ilderton, D. Seipt and B. K\"ampfer.

\end{document}